\documentclass[structabstract]{aa}

\usepackage{amsmath}          
\usepackage{dcolumn}          
\usepackage{natbib}           
\usepackage{graphicx}         
\usepackage{txfonts}          
\usepackage{mathrsfs}         
\usepackage{rotating}         
\usepackage{subfigure}        
\usepackage{lscape}           
\usepackage{afterpage}        
\usepackage{xspace}           

\bibpunct{(}{)}{;}{a}{}{,}    
\newcolumntype{.}{D{.}{.}{-1}}
\newcolumntype{d}[1]{D{.}{.}{#1}}

\newcommand{\mcl}[3]{\multicolumn{#1}{#2}{#3}}
\newcommand{\mrm}[1]{\ensuremath{\mathrm{#1}}}

\providecommand*{\mut}[1]{\ensuremath{\,\mathrm{#1}}}
\providecommand*{\diff}{\ensuremath{\mathrm{d}}}
\providecommand*{\ee}{\ensuremath{\mathrm{e}}}
\providecommand*{\hms}[3]{\ensuremath{{#1}^\mrm{h}{#2}^\mrm{m}{#3}^\mrm{s}}\xspace}
\providecommand*{\dms}[3]{\ensuremath{#1^\circ #2^\prime #3^{\prime\prime}}\xspace}
\usepackage{color}
\usepackage{marginnote}
\begin{document}

\title{Deuterated methanol in the pre-stellar core L1544%
       \thanks{Based on observations carried out with the IRAM 30\,m Telescope. 
               IRAM is supported by INSU/CNRS (France), MPG (Germany), and IGN (Spain).}}

\author{L.~Bizzocchi\inst{1,2} \and P.~Caselli\inst{2} \and S.~Spezzano\inst{3} 
        \and E. Leonardo\inst{1}}

\institute{Centro de Astronomia e Astrof\'isica, Observat\'orio Astron\'omico de Lisboa,
           Tapada da Ajuda, 1349-018 Lisboa (Portugal);
           \email{[bizzocchi,elle]@oal.ul.pt}
           \and
           Centre for Astrochemical Studies,
           Max-Planck-Institut f\"ur Extraterrestrische Physik,
           Gie\ss enbachstra\ss e 1, 85748 Garching (Germany);
           \email{[caselli,bizzocchi]@mpe.mpg.de}
           \and 
           I.Physikalisches Institut, Universtit\"at zu K\"oln
           Z\"ulpicherstar\ss e 77, D-50937 K\"oln (Germany);
           \email{spezzano@ph1.uni-koeln.de}
          }

\titlerunning{CH$_2$DOH in L1544}
\authorrunning{L. Bizzocchi}

\abstract
{High methanol (CH$_3$OH) deuteration has been revealed in Class~0 protostars with the 
 detection of singly, doubly, and even triply D-substituted forms.
 Methanol is believed to form during the pre-collapse phase \emph{via} gas-grain chemistry
 and then eventually injected into the gas when the heating produced by the newly formed
 protostar sublimates the grain mantles.
 The molecular deuterium fraction of the warm gas is thus a relic of the cold pre-stellar 
 era and provides hints of the past history of the protostars.
}
{Pre-stellar cores represent the preceding stages in the process of star formation.
 We aim at measuring methanol deuteration in L1544, a prototypical dense and cold core 
 on the verge of gravitational collapse.
 The aim is to probe the deuterium fractionation process while the ``frozen'' molecular
 reservoir is accumulated onto dust grains.
 }
{Using the IRAM~30\,m telescope, we mapped the methanol emission in the pre-stellar core 
 L1544 and observed singly deuterated methanol (CH$_2$DOH and CH$_3$OD) towards the dust 
 peak of L1544.
 Non-LTE radiative transfer modelling was performed on three CH$_3$OH emissions lines
 at 96.7\,GHz, using a Bonnor--Ebert sphere as a model for the source. 
 We have also assumed a centrally decreasing abundance profile to take the 
 molecule freeze-out in the inner core into
account.
 The column density of CH$_2$DOH was derived assuming LTE excitation and optically 
 thin emission.}
{The CH$_3$OH emission has a highly asymmetric morphology, resembling a non-uniform ring 
 surrounding the dust peak, where CO is mainly frozen onto dust grains.
 The observations provide an accurate measure of methanol deuteration in the cold 
 pre-stellar  gas. 
 The derived abundance ratio is [CH$_2$DOH]/[CH$_3$OH] $= 0.10\pm 0.03$, which is significantly 
 smaller than the ones found in low-mass Class~0 protostars and smaller than the deuterium 
 fraction measured in other molecules towards L1544. 
 The singly-deuterated form CH$_3$OD was not detected at $3\sigma$ sensitivity of 
 7\,mK\,km\,s$^{-1}$, yielding a lower limit of [CH$_2$DOH]/[CH$_3$OD] $\geq 10$, 
 consistent with previous measurements toward Class~0 protostars.
 }
{The low deuterium fractionation observed in L1544 and the morphology of the CH$_3$OH emission
 suggest that we are mainly tracing the outer parts of the core, where CO just started to 
 freeze-out onto dust grains.} 


\maketitle

\section{Introduction} \label{sec:intro}
\indent\indent
Deuterium is known to be \mbox{$\sim 10^{5}$} times less abundant than hydrogen in the 
Universe \citep{Linsky-SSR03-D}; nonetheless, high abundances of D-containing 
isotopologues are common findings in many interstellar environments 
\citep[e.g.,][]{Cecc-PSS02-D,Roueff-SSR03-D}.
During the past couple of decades, observations have revealed large molecular deuteration 
in low-mass pre-stellar cores and Class~0 protostars, where singly, doubly, and --- in 
some instances --- triply deuterated molecules have been detected 
\citep[see, e.g.,][]{Cecc-P&P07-Dchem}.

Methanol (CH$_3$OH) is one of the species that show the highest D-enhancements: 
together with ammonia \citep{Lis-ApJ02-ND3,vdTak-AA02-ND3}, it is one of the two molecules 
for which a triply deuterated form has been observed. 
CD$_3$OH has been revealed in the low-mass protostar IRAS~16293--2422 
\citep{Parise-AA04-CD3OH}, a source showing extreme deuterium enhancements. 
It is, however, not a pathological case, because \citet{Parise-AA06-GrainChem} also found
very high abundances of CH$_2$DOH and CHD$_2$OH in an extended sample of low-mass 
Class~0 protostars.

Theoretical and experimental studies predict that methanol is formed on the grain surface 
by subsequent additions of hydrogen to iced CO \citep{Tielens-AA82-grains,Wata-ApJ02-CH3OH}.
This process is thought to take place during the cold and dense pre-collapse phase 
\citep[e.g.,][]{Oberg-ApJ11-Ice}. 
Later on, the molecule is released in the gas when the heating of newly formed protostar 
sublimates the ices mantles \citep{Cecc-AA01-D2CO}.
Methanol deuteration is thus likely to be produced completely by active grain-surface 
chemistry, controlled by the atomic D content of the accreting gas. 
The high atomic D/H ratio required to account for the observed fractionation 
(0.1--0.2, \citealt{Parise-AA02-CHD2OH}) has been explained by invoking an efficient 
transfer of atomic deuterium from the HD main reservoir \emph{via} the intermediate 
H$_2$D$^+$/D$_2$H$^+$ ions \citep{Roberts-ApJ03-Dfrac,Parise-AA06-GrainChem}, which are very
abundant in the CO-depleted pre-stellar gas 
\citep[e.g.,][]{Caselli-AA03-H2D+,Phil-SFC203-Dmol,Vastel-ApJ04-D2H+,Parise-AA11-D2H+}.
Presently, all the measured methanol deuterations have been satisfactorily reproduced by 
the most recent coupled gas--grain models \citep{Taquet-ApJ12-Deut,Aikawa-ApJ12-Dchem}, 
thus supporting the hypothesis that D-fractionation in methanol is a distinctive relic of 
the protostars' past history.

In this context, it is very interesting to study methanol deuteration in the pre-stellar 
gas. 
Starless cores represent the early stage of low-mass protostar evolution and 
offer the opportunity to probe the initial conditions in the process of star formation.
These objects have a simple structure with no central heating source and little 
(thermal) turbulence, thus providing a favourable environment to study molecular 
deuteration.
In particular, measuring deuterated methanol in pre-stellar cores yields a further 
test of the process responsible for the build-up of D-bearing isotopologue reservoir 
onto grain mantles.

A few studies of methanol in the pre-stellar gas have been reported so far.
Earlier detections of the parent species were accomplished in TMC~1, TMC~1C, 
L134N (= L183), and B335 \citep{Friberg-AA88-CH3OH,Takak-ApJ98-TMC1C,Takak-ApJ00-CH3OH},
and also in some translucent clouds by \citet{Turner-ApJ98-CH3OH}.
Later on, CH$_3$OH was also observed in L1498 and L1517B by \citet{Tafalla-AA06-SCs2}
and, together with its deuterated variant CH$_2$DOH, in the shocked gas of the Class~0 
L1157 source \citep{Codella-ApJ12-L1157}.
Previous detections of deuterated methanol in a sample of pre-stellar cores were also 
reported in a summarised form by \citet{Bacman-MSL07-CH2DOH}, but a detailed analysis of 
these observations is actually missing.

In this paper we report on the observation of CH$_2$DOH towards the starless cloud L1544,
thus providing an accurate assessment of the methanol deuteration in the cold pre-stellar 
gas.
Multi-frequency analysis of the CH$_3$OH emission (including non-LTE modelling) is 
performed in order to derive a reliable value for the column density of the main 
isotopologue.
We finally compare the obtained fractionation ratio with the results derived in Class~0 
protostars and discuss the implications suggested by the predictions based on gas--grain
chemical models.

\section{Observations} \label{sec:obs}
\indent\indent
The methanol data presented here have been collected using the IRAM 30\,m  antenna, 
located at Pico~Veleta (Spain) during three observing sessions in 2012--2013\@.
Single-pointing observations towards the L1544 dust emission peak, located at coordinates
$\alpha$(J2000) = \hms{05}{04}{17.21} and $\delta$(J2000) = \dms{+25}{10}{42.8}
\citep{Caselli-ApJ02-L1544k}, were carried out in October~2012, April~2013, and 
October~2013\@.
The 3\,mm lines of CH$_3$OH were observed using several tunings of the EMIR 
E090 receiver while surveying various organic molecules in L1544 
\citep{Spezz-ApJ13-cC3D2,Spezz-inprep}.
We used the FTS backend in the ``fine'' configuration, resulting in 7.2\,GHz of 
instantaneous bandwidth divided in four sub-bands with a final unsmoothed resolution 
of 50\,kHz.
The CH$_2$DOH transitions were only observed during the April~2013 observing block.
The E090 and E150 receivers were tuned at 89.780 and 134.07\,GHz, respectively, and the
line signals were collected in the lower-inner (LI) sideband.
As for CH$_3$OH, we used the FTS backend in the ``fine'' mode.
Frequency-switching was adopted as observing mode using frequency throws of 3.9\,MHz 
at~3\,mm and 7.8\,MHz at~2\,mm.
The telescope pointing was checked every two hours on nearby bright radio quasars and was 
found accurate to 3--4\arcsec.
Typical system temperatures were 85--130\,K at 97\,GHz and 170--220\,K at 108 and 
134\,GHz\@.
The observed spectra were then converted from the $T^\ast_\mrm{A}$ to the $T_\mrm{mb}$ 
temperature scale adopting $B_\mrm{eff}$ and $F_\mrm{eff}$ values taken from the IRAM
documentation.
The rest frequencies and other spectroscopic parameters of the observed methanol lines 
are reported in Table~\ref{tab:ch3oh_spec}.

L1544 was mapped during the Autumn~2013 session.
We performed a $3\arcmin\times 3\arcmin$ on-the-fly (OTF) map centred on the source dust 
emission peak (see above).
The reference position was set at ($-180\arcsec,180\arcsec$) offset with respect to the 
map centre.
Three methanol lines at 96.7\,GHz were observed using two different E090 setups in the LI 
sideband and FTS in ``fine'' mode. 
The spectral axis was thus sampled with a 50\,kHz channel spacing.
The map area was swept during 4.5\,hours of telescope time by moving the antenna
along an orthogonal pattern of linear paths separated by 8\arcsec\ intervals, corresponding
to roughly one third of the beam FWHM (25.4\arcsec at 96.7\,GHz)\@.
The mapping was carried out in good weather conditions ($\tau\sim 0.03$) and 
a typical system temperature of $T_\mrm{sys}\approx90$\,K\@.
The data processing was done using the GILDAS\footnote%
{See GILDAS home page at the URL:\newline \texttt{http://www.iram.fr/IRAMFR/GILDAS}.} 
software \citep{Pety-SF05-GILDAS}\@.
\begin{table}[h]
  \caption[]{Spectroscopic parameters of the observed methanol lines.
             Data are taken from \citet{Xu-JPC97-CH3OH} for CH$_3$OH, from 
             \citet{Pears-JMS12-CH3OH} for CH$_2$DOH, and from \citet{Ander-ApJS88-CH3OD} 
             for CH$_3$OD}.
  \label{tab:ch3oh_spec}
  \centering\footnotesize
  \begin{tabular}{l d{4} d{3} c  d{6}}
    \hline\hline \noalign{\smallskip}
    Line                              & 
    \mcl{1}{c}{Rest frequency}        &
    \mcl{1}{c}{$E_u/k_b$}             &
    \mcl{1}{c}{$g_u^a$}               &
    \mcl{1}{c}{$A$}                   \\
                                      & 
    \mcl{1}{c}{(MHz)}                 &
    \mcl{1}{c}{(K)}                   &
                                      &
    \mcl{1}{c}{($10^{-5}\,$s$^{-1}$)} \\[0.5ex]
    \hline \noalign{\smallskip}
    CH$_3$OH & & & & \\
    $2_{0,2}-1_{0,1}$, $A^+$      &  96741.375 &  6.96   & 5 & 0.3408 \\
    $2_{1,2}-1_{1,1}$, $E_2$      &  96739.362 & 12.53^b & 5 & 0.2558 \\
    $2_{0,2}-1_{0,1}$, $E_1$      &  96744.550 & 20.08^b & 5 & 0.3407 \\
    $0_{0,0}-1_{1,1}$, $E_1-E_2$  & 108893.963 & 13.12^b & 1 & 1.471  \\
    CH$_2$DOH & & & & \\
    $2_{0,2}-1_{0,1}$, $e_0$      &  89407.817 &  6.40   & 5 & 0.202  \\
    $3_{0,3}-2_{0,2}$, $e_0$      & 134065.381 & 12.83   & 7 & 0.730  \\
    CH$_3$OD  & & & & \\
    $1_{1,0}-1_{0,1}$, $A^--A^+$  & 133925.423 &  8.6    & 3 & 2.982  \\
    \hline \\[-1ex]
    \mcl{5}{l}{$^a$ Rotational degeneracy.} \\
    \mcl{5}{l}{$^b$ Energy relative to the ground $0_{0,0}$, $A$ rotational state.}
  \end{tabular}
\end{table}

\section{Results} \label{sec:results}

\begin{figure*}[tbh]
  \centering
  \includegraphics[angle=0,height=8.0cm]{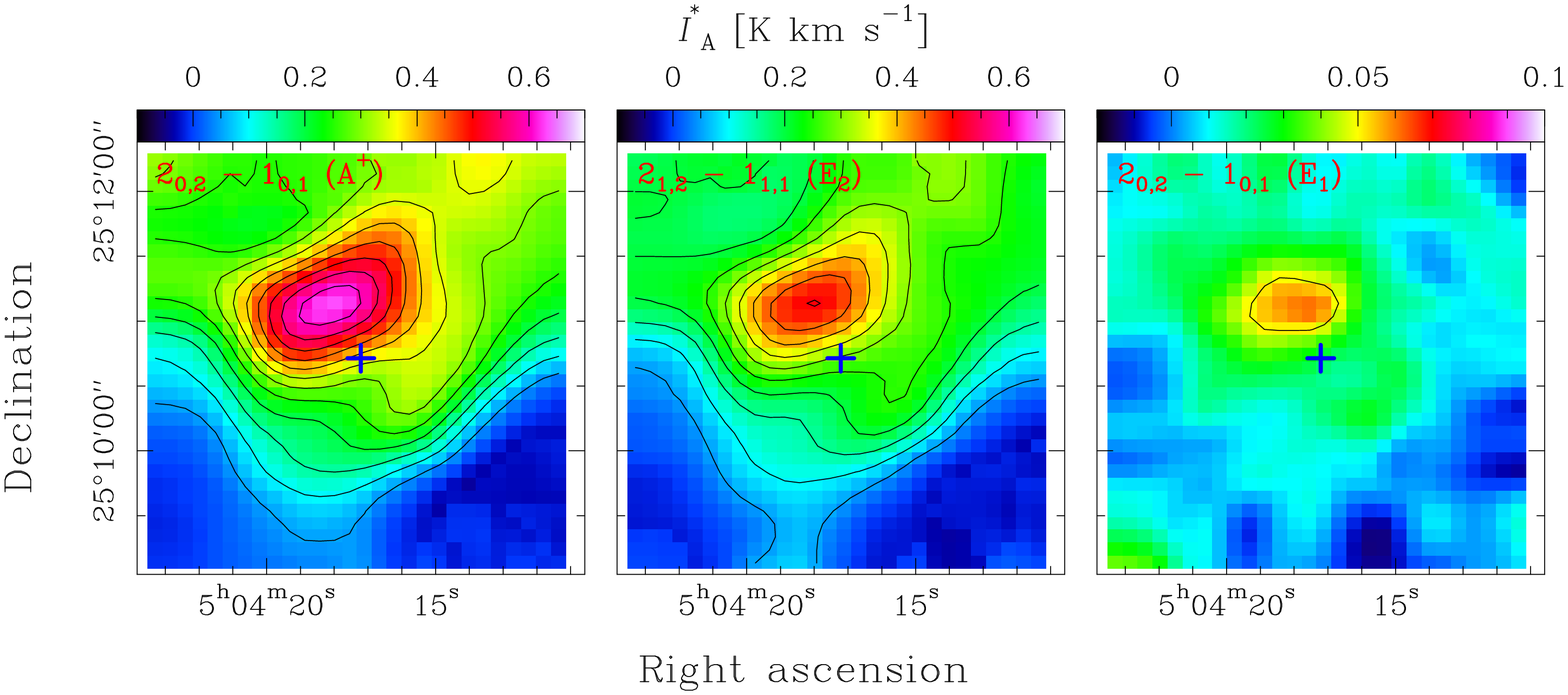} 
  \caption{Intensity maps (units of $T^\ast_\mrm{A}$) of the $2_{0,2}-1_{0,1}$ 
           ($A^+$), $2_{1,2}-1_{1,1}$ ($E_2$), and $2_{0,2}-1_{0,1}$ ($E_1$) transitions of 
           CH$_3$OH integrated over 0.5\,km\,s$^{-1}$ velocity interval\@.
           The L1544 dust peak position located at $\alpha$(J2000) = \hms{05}{04}{17.21}, 
           $\delta$(J2000) = \dms{+25}{10}{42.8} is indicated by the blue cross marker.
           The first contour is at 5$\sigma$ and the increment is 5$\sigma$ for all three
           maps; note however that the colour scale is different for the weak $2_{0,2}-1_{0,1}$ 
           ($E_1$) line (\textit{right panel}).
           The images were smoothed to a 30\arcsec angular resolution to increase the 
           signal-to-noise ratio ($1\sigma\approx 10^{-2}$\,K\,km\,s$^{-1}$).} 
  \label{fig:ch3oh_maps}
\end{figure*}

\begin{figure*}[tbh]
  \centering
  \includegraphics[angle=0,height=7.5cm]{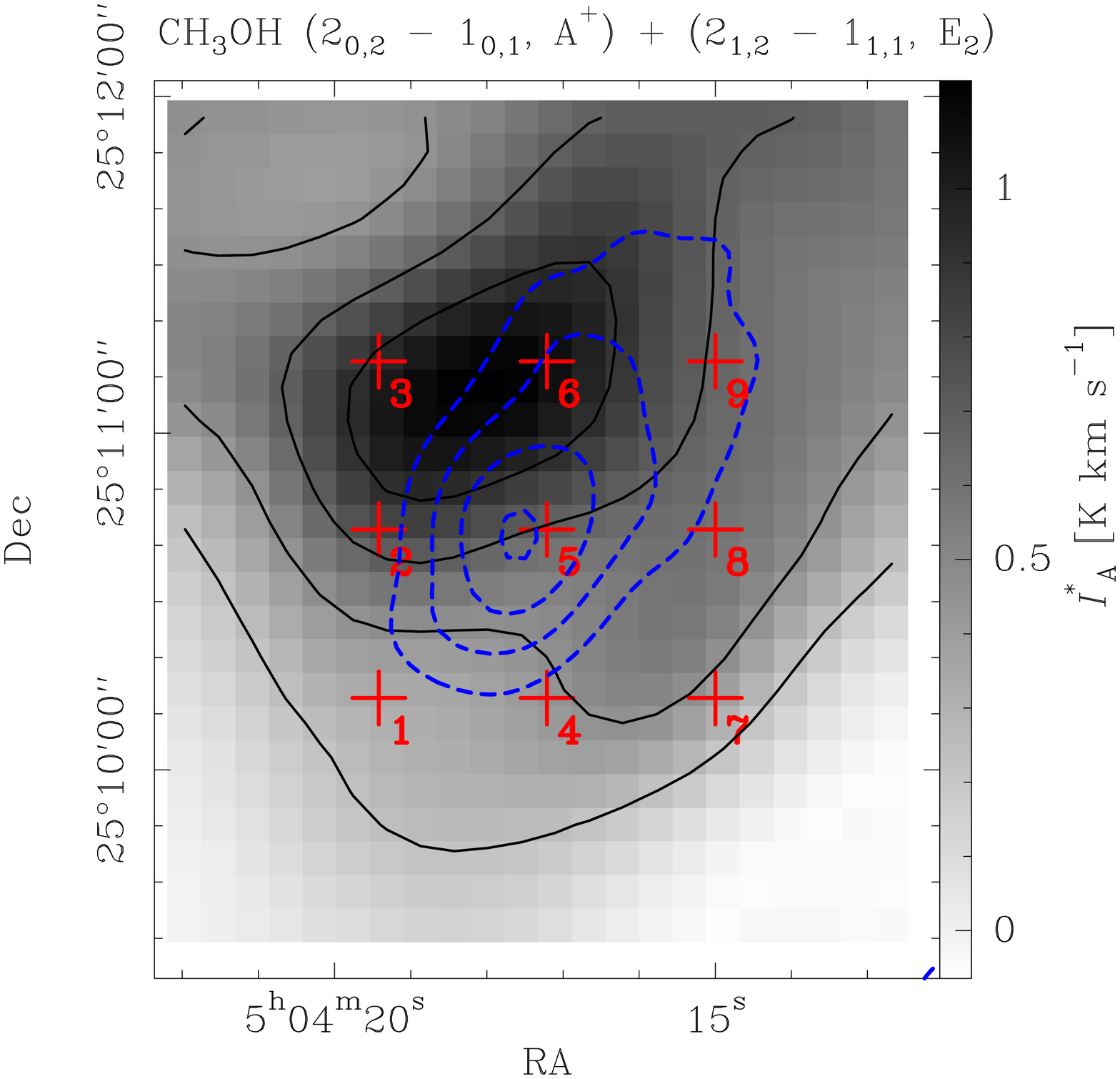} 
  \includegraphics[angle=0,height=7.5cm]{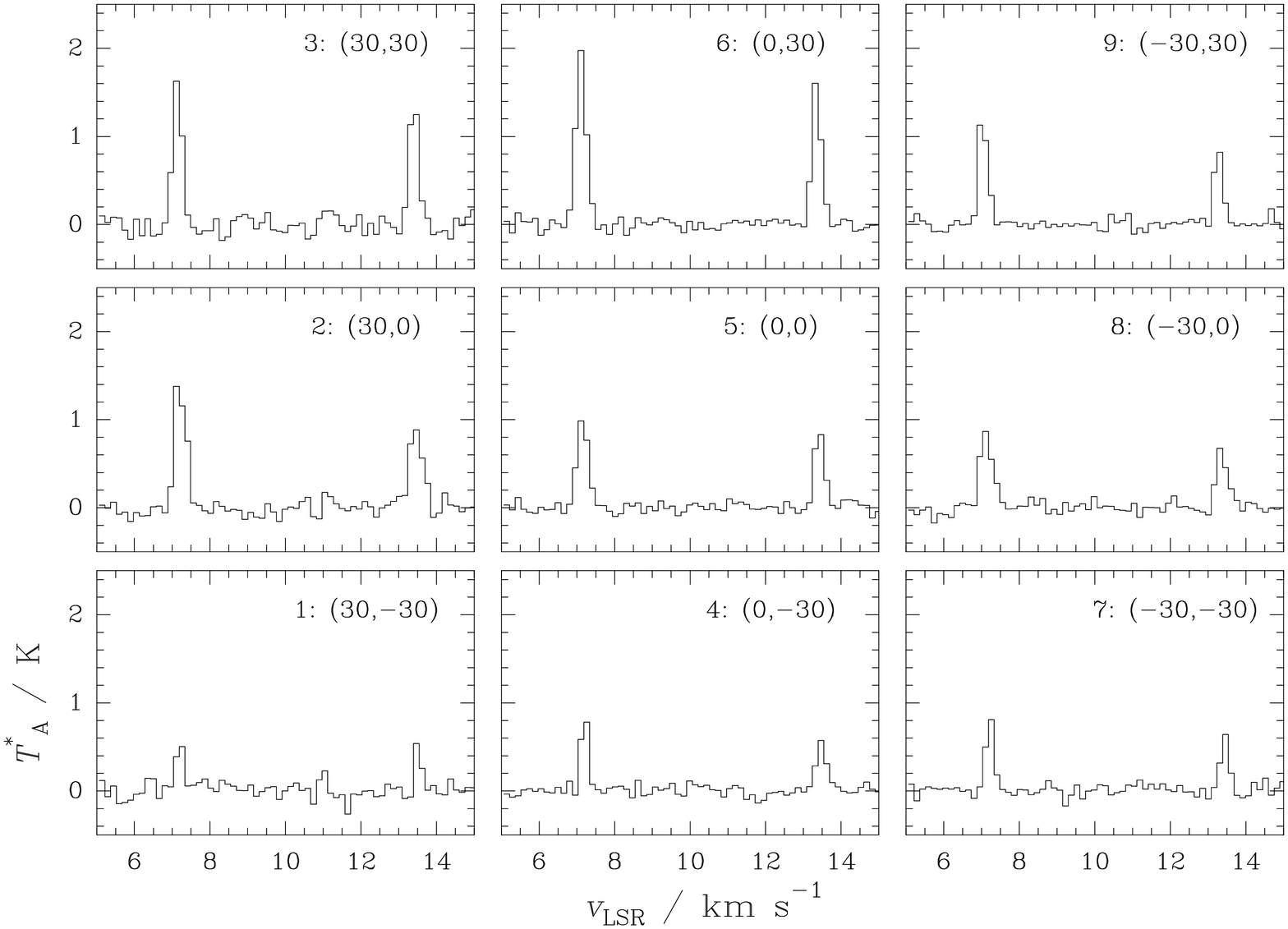}
  \caption{(\textit{Left panel}) Grey-scale map of the summed integrated intensity 
           (units of $T^\ast_\mrm{A}$) of the $2_{0,2}-1_{0,1}$ ($A^+$) and 
           $2_{1,2}-1_{1,1}$ ($E_2$) CH$_3$OH lines (30\arcsec angular resolution).
           Five equally spaced contours from~0.23 to 1.5\,K\,km\,s$^{-1}$ are plotted.
           The blue dashed contours plot the 1.3\,mm continuum intensity map of 
           \citet{WT-MNRAS99-SFIII} smoothed at 22\arcsec to improve the signal-to-noise
           ratio.
           Contours are at 100, 140, 180, and 220\,mJy flux density levels.
           The red crosses represent the offset positions at which the spectra have been
           extracted. ---
           (\textit{Right panel}) Map spectra of the CH$_3$OH transitions towards the nine
           red crosses shown in the left panel.
           The velocity axis is centred on the $2_{0,2}-1_{0,1}$ ($A^+$) line at 
           7.2\,km\,s$^{-1}$.
           The vertical axis of each spectrum represents the $T_\mrm{A}^\ast $ scale in 
           K, as shown in the lower leftmost panel.}
           \label{fig:ch3oh_grid}
\end{figure*}

\begin{figure*}[tbh]
  \centering
  \includegraphics[width=8.5cm]{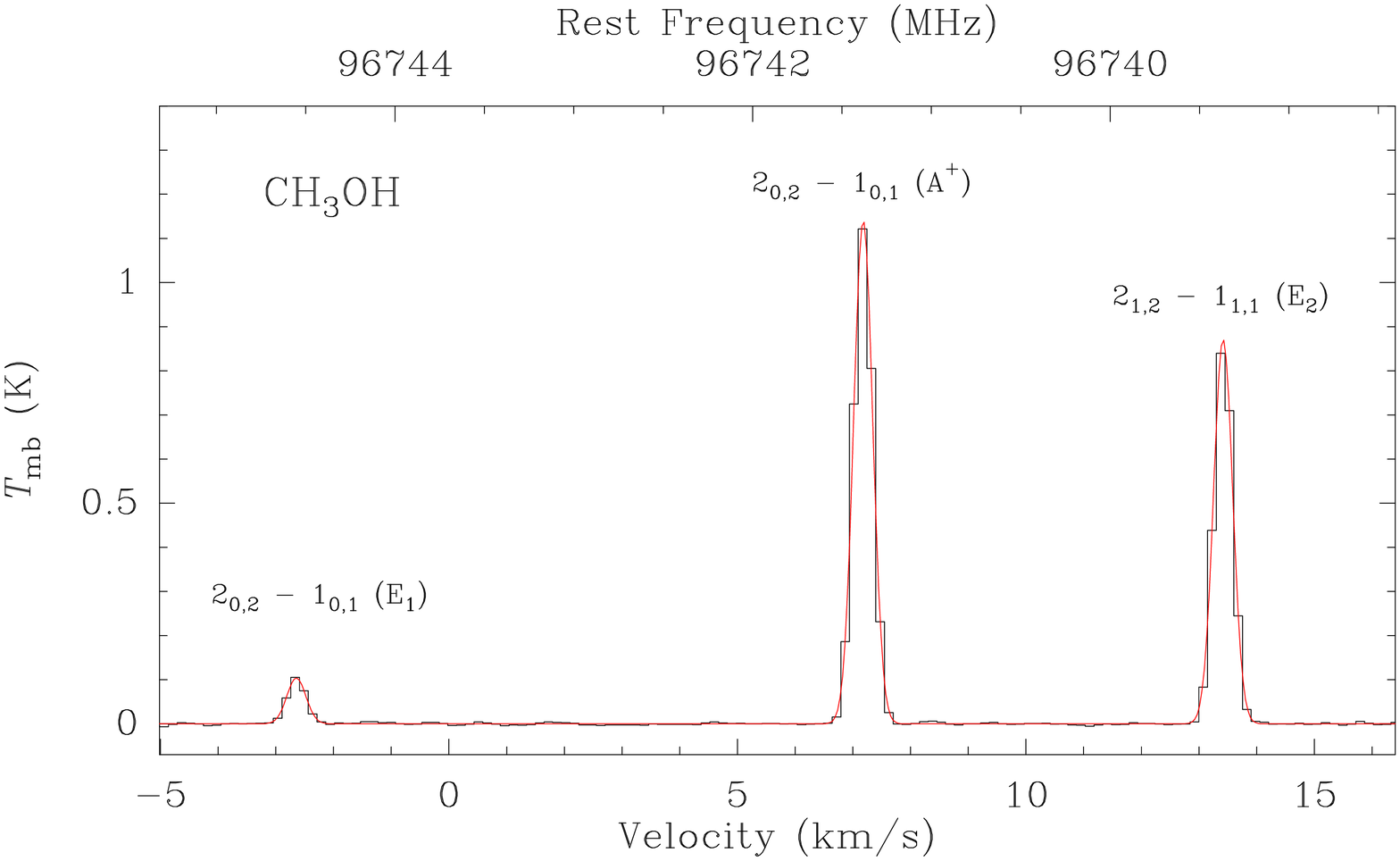}
  \quad
  \includegraphics[width=8.5cm]{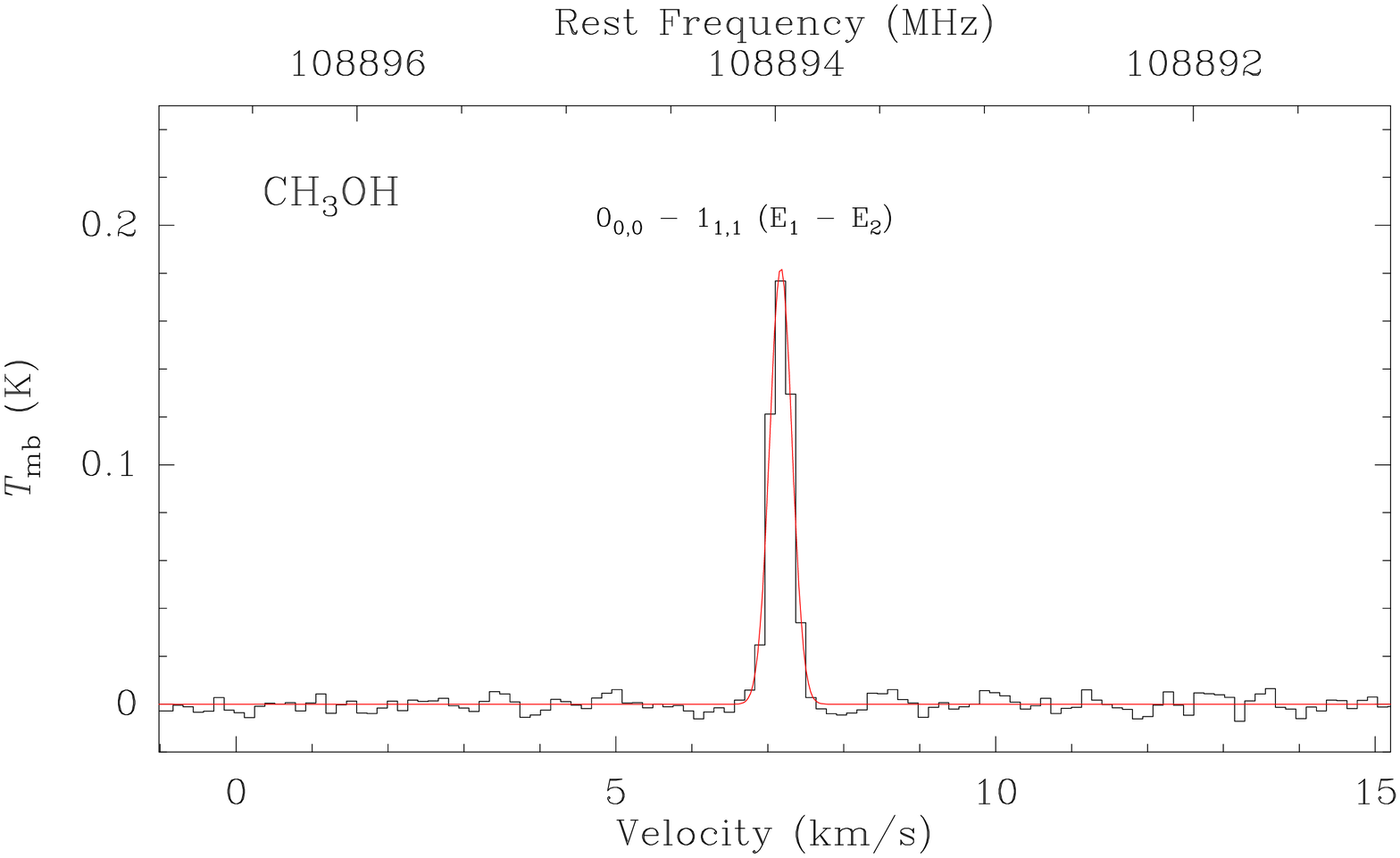}
  \caption{CH$_3$OH lines observed towards the (0\arcsec,0\arcsec) position of L1544\@.
           (\textit{left}) group of $2_{1,2}-1_{1,1}$ and $2_{0,2}-2_{0,1}$ 
           transitions. --- 
           (\textit{right}): $0_{0,0}-1_{1,1}$ ($E_1-E_2$) line\@.
           The spectral \textit{rms} is $\sim$3\,mK\@.
           The red solid line plots the Gaussian fit obtained using CLASS\@.}
  \label{fig:ch3oh_lines}
\end{figure*}

\subsection{Mapping of CH$_3$OH} \label{sec:maps}
\indent\indent
Figure~\ref{fig:ch3oh_maps} shows the maps obtained for the CH$_3$OH 
$2_{0,2}-1_{0,1}$ ($A^+$), $2_{1,2}-1_{1,1}$ ($E_2$), and $2_{0,2}-1_{0,1}$ ($E_1$)
transitions. 
The last is appreciably weaker because it comes from a level of higher energy 
(see Table~\ref{tab:ch3oh_spec}).
The reference centre of the map is indicated and coincides with the maximum of the 
continuum 1.3\,mm emission \citep{WT-MNRAS99-SFIII}.

In the right-hand panel of Figure~\ref{fig:ch3oh_grid}, we plot a grid of spectra of the 
$2_{0,2}-1_{0,1}$ ($A^+$) and $2_{1,2}-1_{1,1}$ ($E_2$) CH$_3$OH lines.
They were taken in nine positions separated by 30\arcsec, as shown in the left-hand 
panel of the figure.
Here, the grey scale represents the summed integrated intensity of the two lines, whereas 
the blue contours plot the 1.3\,mm continuum emission map \citep{WT-MNRAS99-SFIII} smoothed
at 22\arcsec\@.
The lines peak strongly at the position~6, (0\arcsec, 30\arcsec), whereas the corresponding
signals are reduced by one-half at the position~5, which coincides with the map reference 
centre and is also very close to the maximum of the dust emission.

The CH$_3$OH emission thus differs from the dust continuum and presents a single 
peak offset to the north-east.
A weak intensity enhancement located to the south-west is also discernible, thus suggesting 
the presence of a highly asymmetrical broken ring distribution and reflecting the presence 
of a central CO depletion hole. 
This feature is apparent in the C$^{17}$O integrated intensity map shown in 
\citet{Caselli-ApJ99-COdep} (see their figure~1).
Also, the observed azimuthal asymmetry of the methanol emission is likely to be linked to 
slight inhomogeneities of the CO depletion, owing to the non-spherical and cometary-shaped
morphology \citep[e.g.,][]{Tafalla-AA04-SCs,Crapsi-AA07-L1544}.

\subsection{CH$_3$OH single-pointing observations} \label{sec:ch3oh_pt}
\indent\indent
Besides mapping, single-pointing, sensitive observations were performed towards the L1544
centre, corresponding to position~5 of Figure~\ref{fig:ch3oh_grid} (dust emission
peak).
Four CH$_3$OH emission lines were observed: three falling in a small frequency interval 
at 96.7\,GHz (the same mapped, see Figure~\ref{fig:ch3oh_maps}), plus one line at 
108.9\,GHz\@.
The observations are shown in Figure~\ref{fig:ch3oh_lines}.
Line profiles were analysed using the GAUSS task of CLASS and the resulting data are 
reported in Table~\ref{tab:ch3oh_fits}\@.
The fit of the three closely spaced lines $2_{0,2}-1_{0,1}$ ($A^+$),
$2_{1,2}-1_{1,1}$ ($E_2$), and $2_{0,2}-1_{0,1}$ ($E_1$) was carried out by adjusting 
only one ``common'' line width.
Given the relatively small number of channels used in the least-squares fit, the 
statistical error on the optimised parameters yielded by the GAUSS procedure could be 
optimistic.
Thus, to be on the safe side, we conservatively quote 
$3\sigma$ uncertainties in Table~\ref{tab:ch3oh_fits}.

\begin{figure}[tbh]
  \centering
  \includegraphics[width=8.5cm]{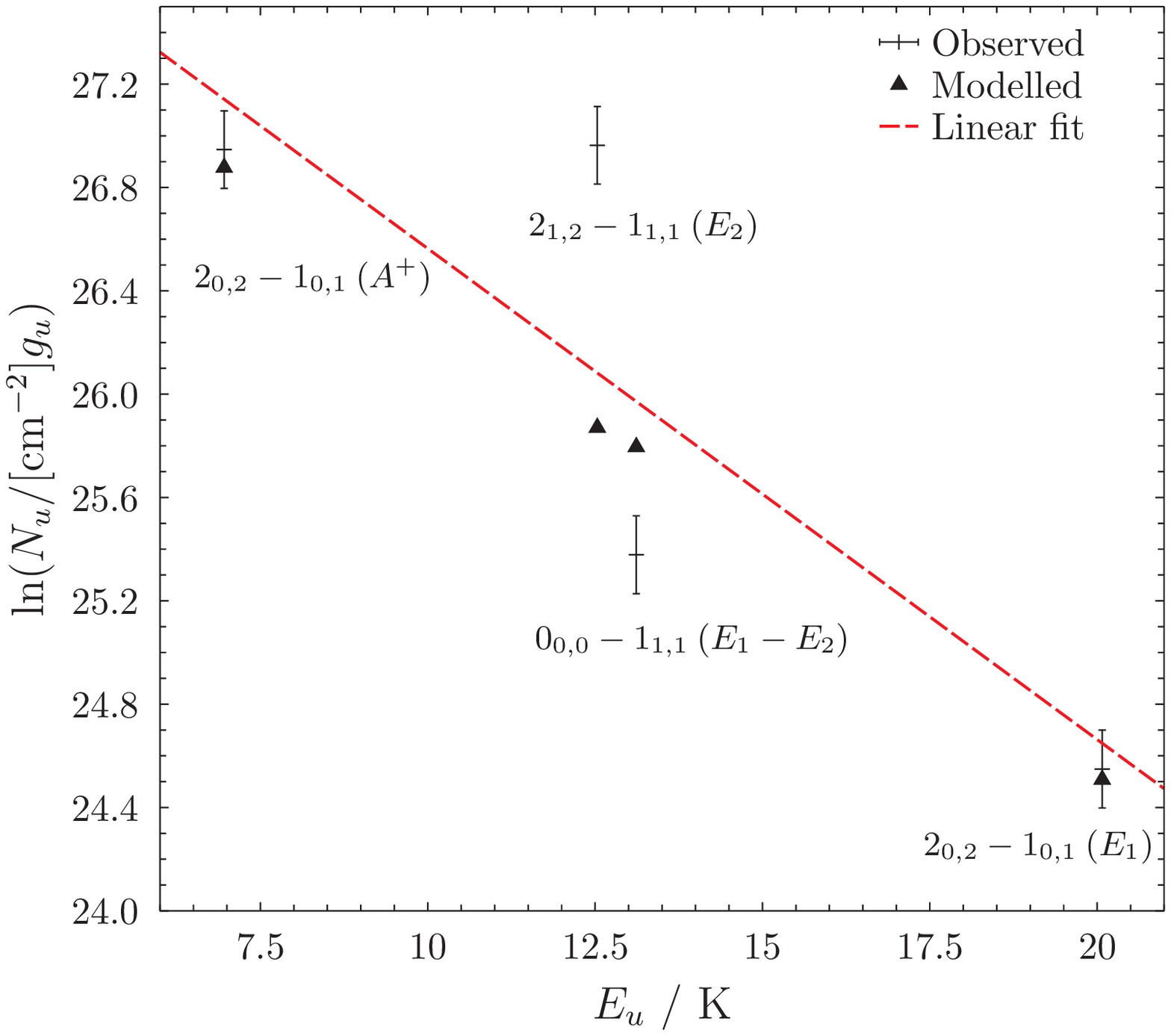}
  \caption{Rotational diagram produced by the four CH$_3$OH lines observed towards L1544\@.
           A considerable scatter is apparent.
           The red dashed line represent the ``best'' linear fit of all data and yields
           $T_\mrm{ex} = 6\pm 3$\,K (see text).}
  \label{fig:ch3oh_rotdiag}
\end{figure}

The emitting levels have energies ranging from~7 to~20\,K, thus one may try to derive 
the CH$_3$OH column density, $N$, and the average excitation temperature, $T_\mrm{ex}$, 
through the population diagram method \citep{Goldsm-ApJ99-RotDiag}.
We used here the modified method described by \citet{Numm-ApJS00-RotDiag}, which
also includes the cosmic background emission and the peak optical depth in addition to
$T_\mrm{ex}$ and $N$\@.
In this approach, the optimised parameters were sought by minimising the squared sum of
the error weighted differences between observed and modelled line intensities.
These intensities are derived through the radiative transfer equality
\begin{equation} \label{eq:Tmb}
 T_\mrm{mb} = \eta_\mrm{bf}\,[J_\nu(T_\mrm{ex}) - J_\nu(T_\mrm{bg})]\,(1 - \ee^{-\tau_\nu}) \,,
\end{equation}
where $\eta_\mrm{bf}$ is the source-beam filling factor, and $J_\nu(T)$ is the equivalent 
Rayleigh-Jeans temperature. 
Assuming a Gaussian line profile, the peak opacity of each transition is obtained as
\begin{equation} \label{eq:tau}
 \tau_\nu^\mrm{(peak)} = 
            \sqrt{\frac{\ln 2}{16\pi^3}}\frac{c^3 A g_u N}{\nu^3 Q(T_\mrm{ex})\Delta v}
            \ee^{-E_u/kT_\mrm{ex}}\,\left(\ee^{h\nu/kT_\mrm{ex}} - 1\right) \,.
\end{equation}
Here, $\nu$ is the emission frequency, $\Delta v$  the FWHM line width in units of 
velocity, $E_u$  the energy of the upper level (as listed in Table~\ref{tab:ch3oh_spec}), 
$g_u$  the rotational degeneracy, $A$ the Einstein's coefficient for spontaneous emission, 
and  $Q(T)$ is the rotation partition function at temperature $T$\@. 
This is computed by summing over all ($A + E$) rotational levels \citep{Xu-JPC97-CH3OH}, 
whose energies are available at the CDMS\footnote{
  Cologne Databases for Molecular Spectroscopy, URL: \newline
  \texttt{http://www.astro.uni-koeln.de/cdms/}} \citep{Muller-JMS05-CDMS}.
Throughout the calculations, the average FWHM line width of 0.37\,km\,s$^{-1}$ was used 
and the beam filling factor $\eta_\mrm{bf}$ was set to unity.

\begin{table}[tbh]
  \caption[]{Results of the CLASS GAUSS fit on the observed spectral profile of the
             methanol lines observed towards L1544.
             Numbers in parentheses refer to $3\sigma$ uncertainties in units of the 
             last quoted digit.}
  \label{tab:ch3oh_fits}
  \centering\footnotesize
  \begin{tabular}{l d{8} d{6} d{8}}
    \hline\hline \noalign{\smallskip}
    Line   & 
    \mcl{1}{c}{$v_{LSR}$}                 &
    \mcl{1}{c}{$\int T^\ast_\mrm{A}\,v$}  &
    \mcl{1}{c}{$\Delta v$}                \\
           & 
    \mcl{1}{c}{(km$\,$s$^{-1}$)}          &
    \mcl{1}{c}{(mK$\,$km$\,$s$^{-1}$)}    &
    \mcl{1}{c}{(km$\,$s$^{-1}$)}          \\[0.5ex]
    \hline \noalign{\smallskip}
    CH$_3$OH & & & \\
    $2_{0,2}-1_{0,1}$ ($A^+$)     &  7.1804(11) & 472.4(25) & 0.3888(12)^a \\
    $2_{1,2}-1_{1,1}$ ($E_2$)     &  7.1808(14) & 360.7(25) & 0.3888(12)^a \\
    $2_{0,2}-1_{0,1}$ ($E_1$)     &  7.194(11)  &  43.0(20) & 0.3888(12)^a \\
    $0_{0,0}-1_{1,1}$ ($E_1-E_2$) &  7.1723(78) &  67.1(30) & 0.346(17)    \\
    CH$_2$DOH & & & \\
    ($2_{0,2}-1_{0,1}$, $e_0$)    &  6.889(24)  &  30.5(42)  & 0.358(60)   \\
    ($3_{0,3}-2_{0,2}$, $e_0$)    &  6.939(22)  &  31.3(45)  & 0.313(48)   \\
    \hline \\[-1ex]
    \mcl{4}{l}{$^a$ Fitted as average value.}
  \end{tabular}
\end{table}

Once the best-fit $N$ and $T_\mrm{ex}$ are determined through Eqs.~\eqref{eq:Tmb} 
and~\eqref{eq:tau}, the match between observed and modelled data can be presented in a 
population diagram fashion, $\ln (N_u/g_u)$ \textit{vs.}\ $E_u$, where the $N_u$ value 
for each transition is derived from the corresponding peak optical opacity 
through
\begin{equation} \label{eq:Nu}
 N_u = [B_\nu(T_\mrm{ex}) - B_\nu(T_\mrm{bg})] \sqrt{\frac{4\pi^3}{\ln 2}} 
        \:\frac{\tau_\nu^\mrm{(peak)}\Delta v}{hcA} \,.
\end{equation}
The result of this comparison is illustrated in Figure~\ref{fig:ch3oh_rotdiag}\@.
The observed points exhibit a considerable scatter, largely exceeding the estimated error 
bars ($\sim$15\%, including calibration and pointing uncertainties).
The largest deviations are shown by the $2_{1,2}-1_{1,1}$ ($E_2$) line, which appears
substantially brighter than expected.
The best-fit modelled points instead lie on a straight line, not far from the linear 
fit obtained under the assumptions of optically thin emission and negligible background 
radiation (\citealt{Goldsm-ApJ99-RotDiag}).
Indeed, the derived peak optical depths are moderate ($\tau < 0.4$) and the cosmic 
background ($T_\mrm{bg} = 2.7$\,K) emission merely acts as an offset.
Figure~\ref{fig:ch3oh_rotdiag} clearly indicates that excitation anomalies are present,
thus preventing determination of a unique excitation temperature for all the
rotational levels involved in the observed emissions.
Indeed, the analysis yielded poorly constrained results: $T_\mrm{ex} = 6\pm 3$\,K and a 
column density value $N = (1.9\pm 1.9)\times 10^{13}$\,cm$^{-2}$\@.

\subsection{Non-LTE modelling} \label{sec:nlte}
\indent\indent
To achieve a better constraint for the CH$_3$OH column density in L1544, 
we performed a non-LTE modelling using the radiative transfer code MOLLIE 
\citep{Keto-APJ10-HypRT} and the L1544 physical model with central density of 
$1\times 10^7$\,cm$^{-3}$ described in \citet{Keto-MNRAS14-H2O}.
\begin{figure}[b]
  \centering
  \includegraphics[width=8.5cm]{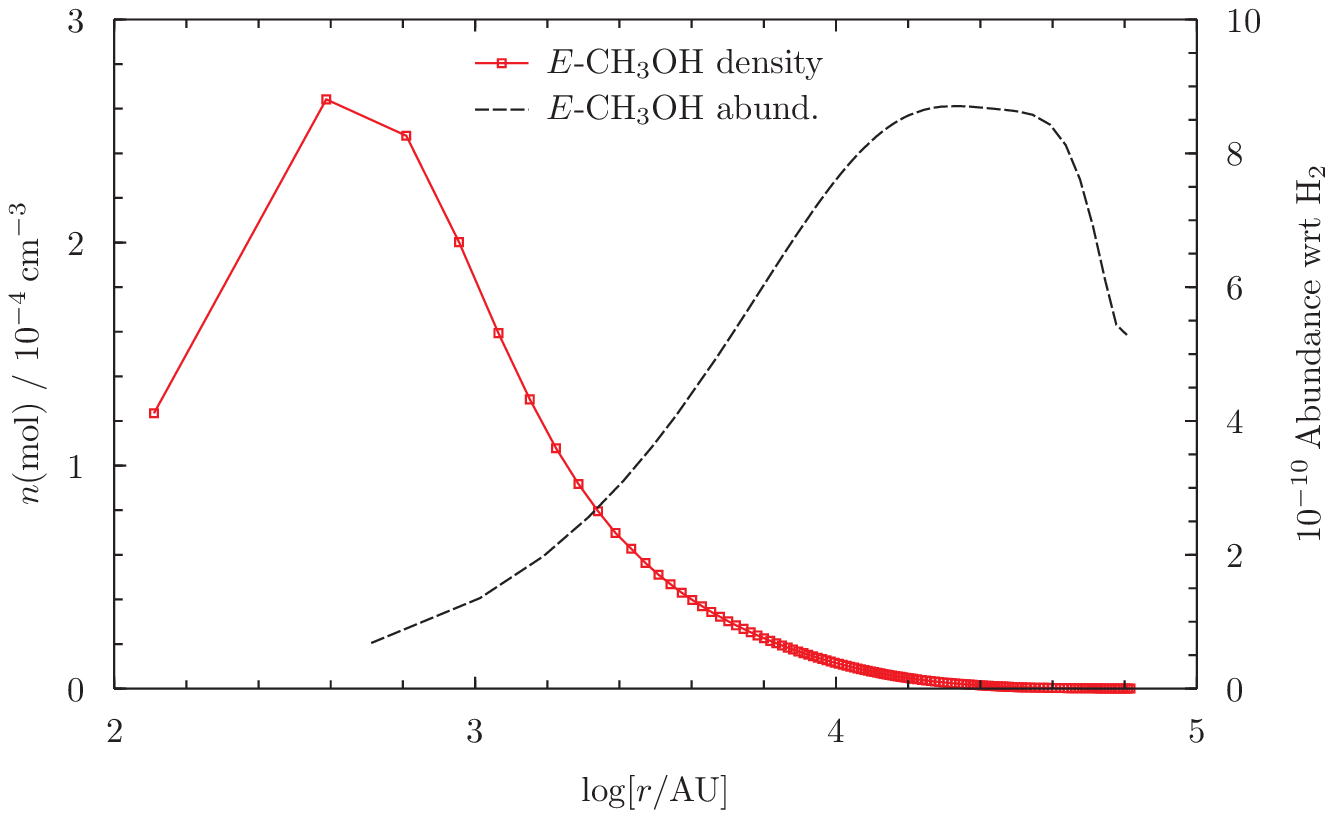}
  \caption{(\textit{Left panel}) Radial profiles for the $E$-CH$_3$OH abundance (black) 
           and density (red) corresponding to the best-fit input abundance of
           $9.2\times 10^{-10}$. 
           The radial trends were produced by the simple chemistry model implemented in 
           the radiative transfer code MOLLIE (see text).}
  \label{fig:Mollie_prof}
\end{figure}
The model is computed with a spherical Lagrangian hydrodynamic code with the gas 
temperature set by radiative equilibrium between heating by external starlight
and cosmic rays and cooling by molecular line and dust radiation (see 
\citealt{Keto-ApJ08-SC} for a comprehensive discussion of the theory). 
Even if the L1544 emission maps show that the cloud is not spherical but instead has an 
elongated shape, the adopted model is simple, physically motivated and certainly adequate 
to model observation data averaged over a single dish beam profile.
For the statistical equilibrium calculation, we used de-excitation rates for 
$p$-H$_2$/$A$-CH$_3$OH and $p$-H$_2$/$E$-CH$_3$OH collisional systems 
\citep{Rabli-MNRAS10-CH3OH_H2} available at the LAMDA database \citep{Schoi-AA05-LAMDA}.
Collisional data for $o$-H$_2$ are lacking, but this does not represent a problem for 
our modelling because the H$_2$ \textit{ortho}-to-\textit{para} ratio (OPR) is expected to 
be very low in pre-stellar cores \citep[see, e.g.,][]{Walmsley-AA04-depl,Sipila-AA13-HDdep}.
Simulations were run separately for $A$ and $E$ symmetry species of methanol 
and considering only rotational levels below 36\,K\@.

MOLLIE implements a simple CO chemistry \citep{Keto-ApJ08-SC} to take into account both 
the molecule freeze-out towards the inner cloud core and the photo-dissociation due to 
the UV stellar field at the edges.
Since CO is considered to be the main precursors of methanol, we used the same model to 
describe the molecular abundance trend across the core.
The ``nominal'' CH$_3$OH abundance, provided as the input parameter to MOLLIE, is thus
internally translated in a radial abundance profile.
As an example, the $E$-CH$_3$OH radial abundance and density profile corresponding to the 
best-fit input abundance (see below) are illustrated in Figure~\ref{fig:Mollie_prof}.
The resulting beam-averaged value of the CH$_3$OH column density (including both
$A$ and $E$ species) is $2.66\times 10^{13}$\,cm$^{-2}$\@.

At the end of the computation, the code outputs a data cube representing the spectral 
distribution of the emerging radiation field.
After convolution with the appropriate telescope beam (25.5\arcsec at 97.6\,GHz, 
22.9\arcsec at 108.9\,GHz), the modelled spectra were extracted from the central pixel 
and compared to the observations.
\begin{figure}[h]
  \centering
  \includegraphics[width=8.5cm]{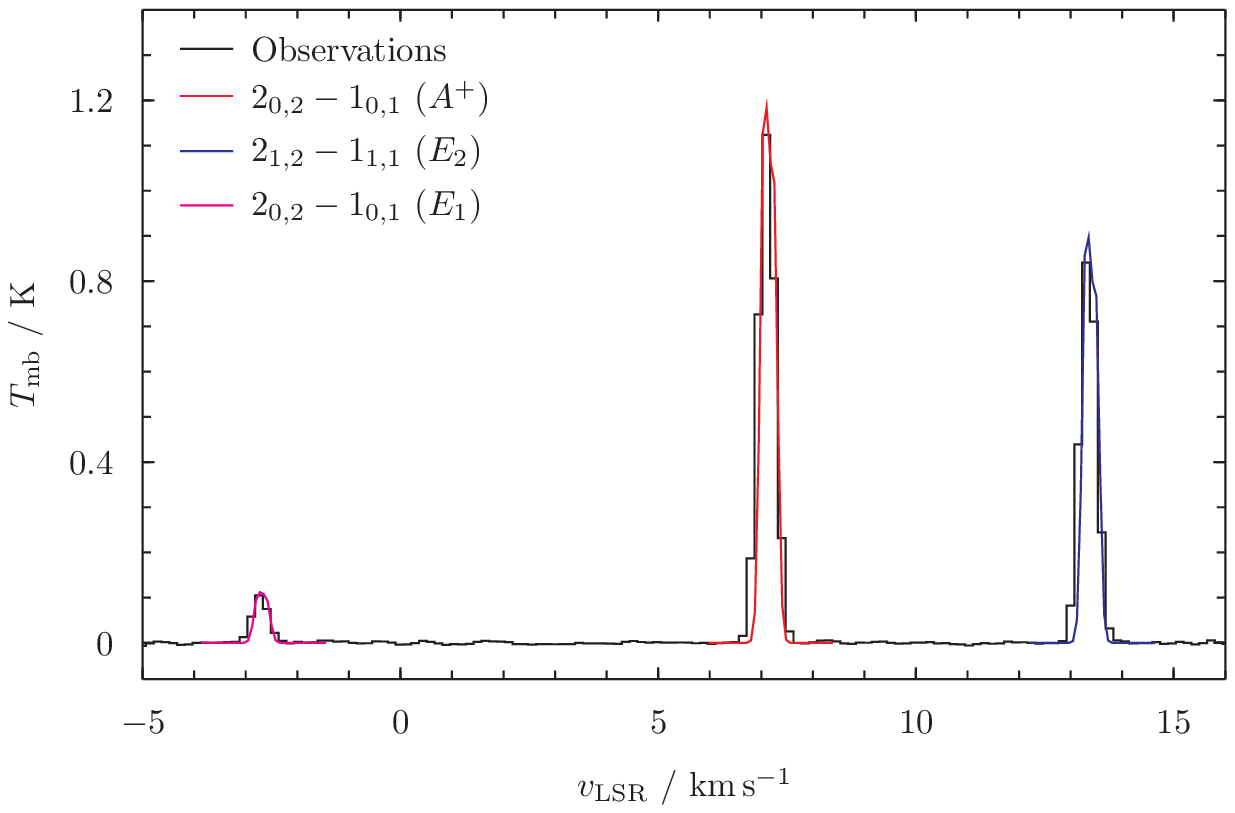}
  \caption{Observed \textit{vs.}\ modelled spectrum of CH$_3$OH 
           in L1544.
           The black histogram shows the observations.
           Coloured lines indicate the best-fit modelled spectra.
           The velocity axis is centred on the $2_{0,2}-1_{0,1}$ ($A^+$) line at 
           7.2\,km\,s$^{-1}$.}
  \label{fig:Mollie_mod}
\end{figure}
Optimal $\chi^2$-fit between observed and modelled spectral profiles were found 
for the $2_{0,2}-1_{0,1}$ ($A^+$), $2_{1,2}-1_{1,1}$ ($E_2$), and 
$2_{0,2}-1_{0,1}$ ($E_1$) transitions using input abundances of $9.5\times 10^{-10}$ and 
$9.2\times 10^{-10}$, for $A$ and $E$ species, respectively. 
The modelling results for these three emissions falling at 96.7\,GHz are shown in 
Figure~\ref{fig:Mollie_mod}.

Despite the success in reproducing these three lines, the fit underestimates the brightness 
of the observed $0_{0,0}-1_{1,1}$ ($E_1-E_2$) emission at 108.9\,GHz
by a factor of~4.
\begin{figure}[tbh]
  \centering
  \includegraphics[width=8.5cm]{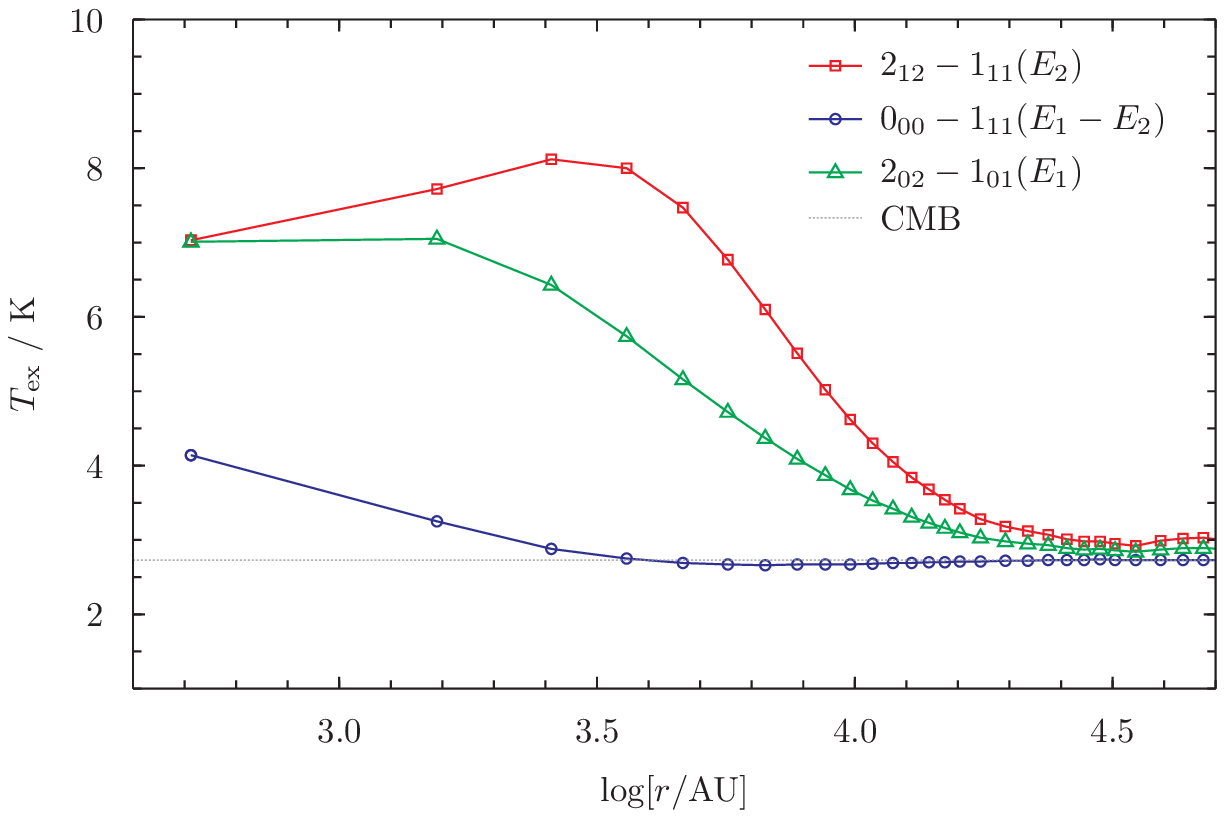}
  \caption{Excitation temperature \emph{vs}.\ cloud radius plot for the 
           $2_{1,2}-1_{1,1}$, $2_{0,2}-1_{0,1}$, and $0_{0,0}-1_{1,1}$ transitions of 
           $E$-CH$_3$OH as computed by the present best-fit radiative transfer 
           modelling.}
  \label{fig:Mollie_Tex}
\end{figure}
Figure~\ref{fig:Mollie_Tex} illustrates the trend towards the excitation temperatures for 
the modelled $E$-CH$_3$OH transitions as a function of the cloud radius. 
It shows that the $0_{0,0}-1_{1,1}$ line is sub-thermally excited even at the high gas 
density of the cloud centre, where the other two $E$ lines are instead thermalised.
The reason for this behaviour is easily understood by inspecting Figure~\ref{fig:E-levs},
which shows the bottom part of the rotational level diagram of $E$-CH$_3$OH, together with 
the radiative and collisional transitions considered in the present modelling.
The $2_{0,2}$ and $2_{1,2}$ upper levels are connected to their corresponding lower 
state within the $K_a = 0$ or $K_a = 1$ manifolds by ``strong'' collisional transitions
(i.e., upward rate $>10^{-11}$\,cm$^3$\,s$^{-1}$ at 10\,K), and other weaker 
$E_1-E_2$ connections also exist. 
On the other hand, the collisional transitions connecting $0_{0,0}$ to the lower $1_{1,1}$ 
and $2_{1,2}$ states have rate coefficients that are set to zero in the present 
$E$-CH$_3$OH/$p$-H$_2$ data set \citep{Rabli-MNRAS10-CH3OH_H2}.
As discussed in \citet{Rabli-MNRAS10-CH3OH_He}, this is an artefact of the coupled state
(CS) approximation used in the production of the collision cross sections, and it is very 
likely that the collisional $0_{0,0}-1_{1,1}$ and $0_{0,0}-2_{1,2}$ transitions actually 
have small, but non-zero, rate coefficients. 

Repeated tests using altered sets of collisional data showed that rate coefficients as 
small as $3\times 10^{-11}$\,cm$^3$\,s$^{-1}$ (\emph{ca.}\ one third of the 
$2_{1,2}-1_{1,1}$ upward rate) allow for a satisfactory modelling of the 
$0_{0,0}-1_{1,1}$ ($E_1-E_2$) emission without significantly altering the fit quality of 
the $2_{1,2}-1_{1,1}$ ($E_2$) and $2_{0,2}-1_{0,1}$ ($E_1$) transitions.
This finding thus suggests that the difficulties encountered in modelling the 
$0_{0,0}-1_{1,1}$ ($E_1-E_2$) line are indeed caused by small inaccuracies in the
collisional dataset used.
\begin{figure}[tbh]
  \centering
  \includegraphics[width=8.5cm]{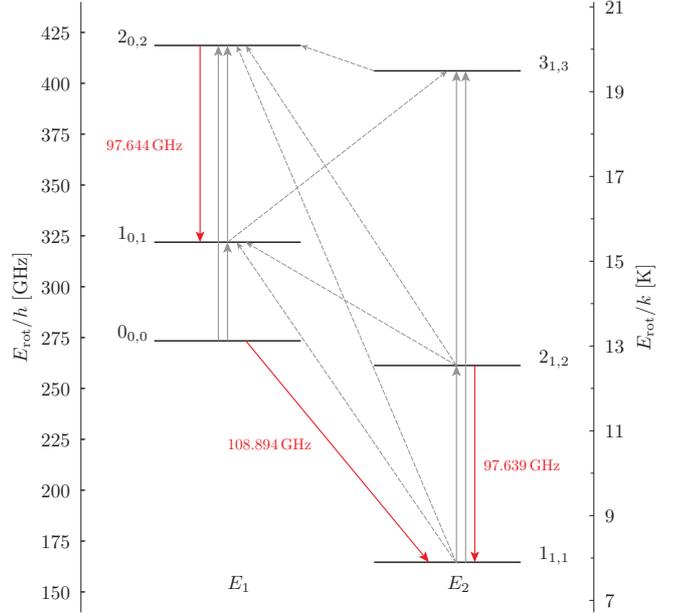}
  \caption{Rotational energy plot for the lowest levels of $E$-CH$_3$OH.
           Active collisional channels are indicated by the upward grey arrows, with solid 
           lines marking ``strong'' collisional transitions (upward rate 
           $>10^{-11}$\,cm$^3$\,s$^{-1}$ at 10\,K, \citealt{Rabli-MNRAS10-CH3OH_H2}).
           Red solid arrows      indicate the \emph{observed} radiative transitions.}
  \label{fig:E-levs}
\end{figure}


An estimate of the error bar associated to the CH$_3$OH column density was obtained as in 
\citet{Bizz-AA13-L1544}.
The Gaussian width of the $\chi^2$ profile (plotted \textit{vs.}\ the free parameter) 
is used to get an estimate of the uncertainty involved in the optimisation process 
($\sim 19\%$).
A further 10\% error is added in quadrature to take the telescope calibration into account, 
yielding a final estimate of a 22\% relative uncertainty.
We thus ended with \mbox{$N$(CH$_3$OH) = $(2.7\pm 0.6)\times 10^{13}$\,cm$^{-2}$}.
This value is not far from the uncertain results derived using an LTE approach in 
\S~\ref{sec:ch3oh_pt}, and it provides a more accurate constraint for the CH$_3$OH 
column density in L1544\@.

\subsection{Singly deuterated methanol (CH$_2$DOH)} \label{sec:ch2doh}
\begin{figure*}[t]
  \centering
  \includegraphics[width=8.5cm]{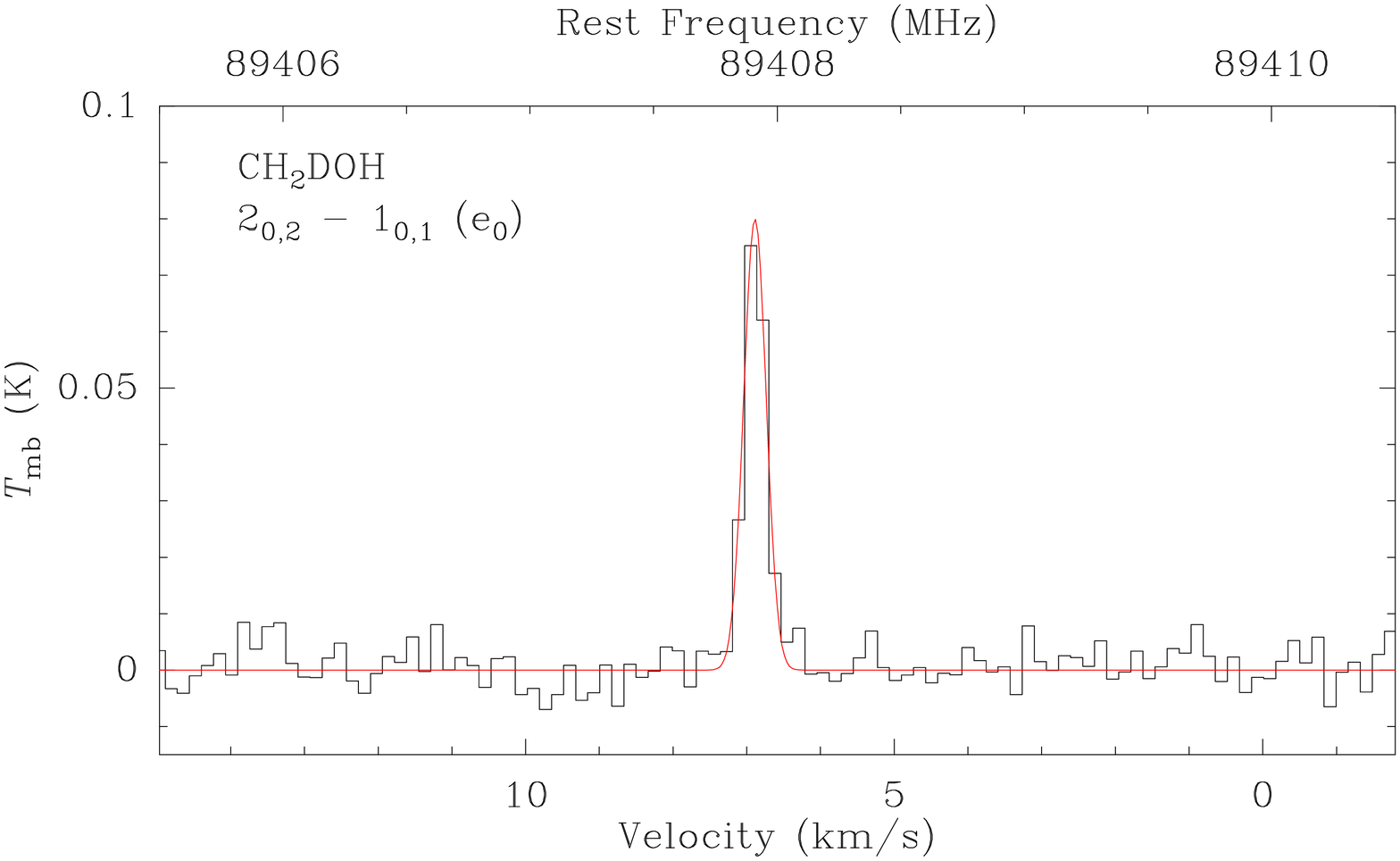}
  \quad
  \includegraphics[width=8.5cm]{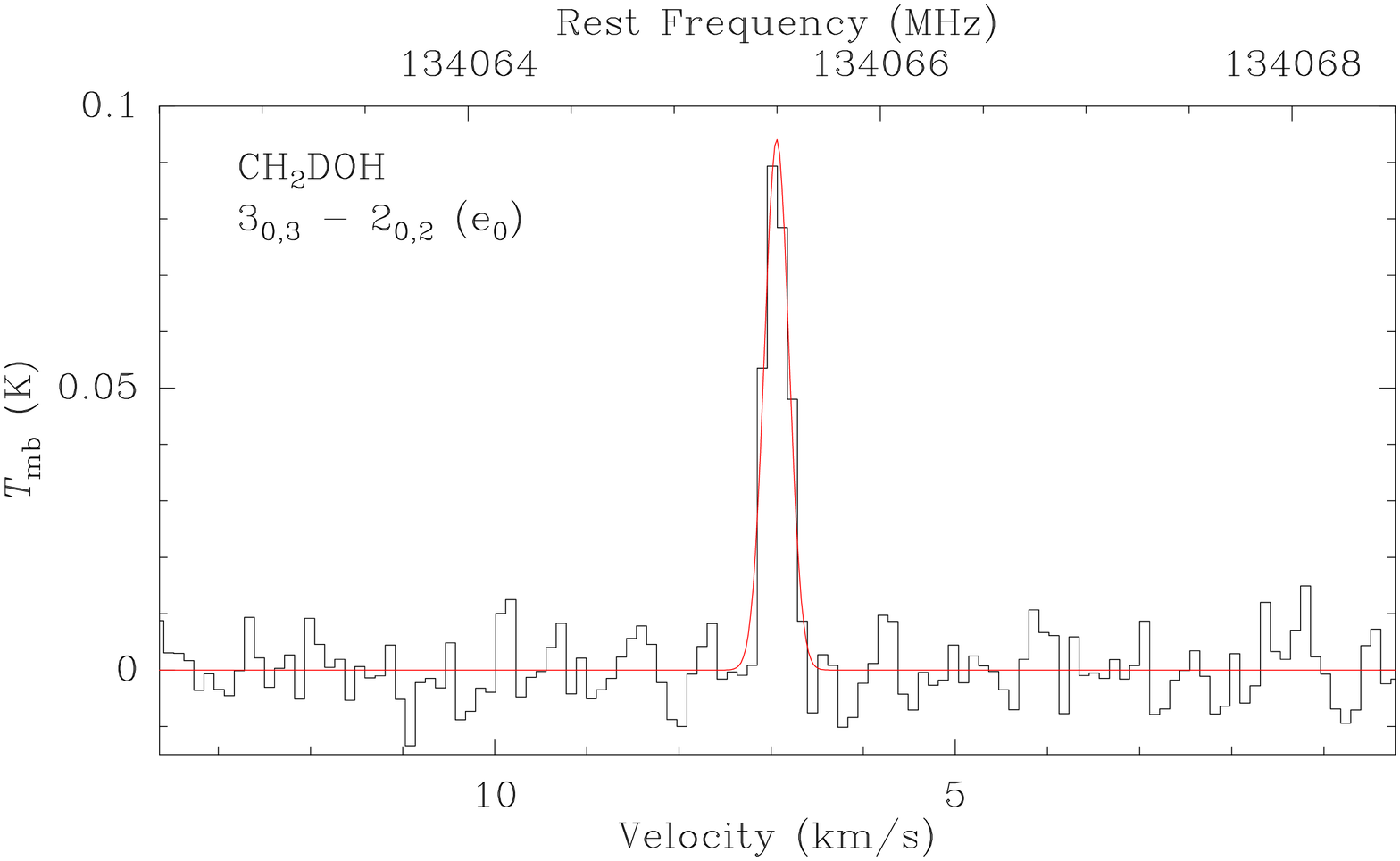}
  \caption{CH$_2$DOH lines observed towards the (0\arcsec,0\arcsec) position of L1544\@.
           (\textit{left}) $2_{0,2}-1_{0,1}$ ($e_0$) ---
           (\textit{right}): $3_{0,3}-2_{0,2}$ ($e_0$)\@.
           The spectral \textit{rms} is $\sim$5\,mK\@.
           The red solid line plots the Gaussian fit obtained using CLASS\@.}
  \label{fig:ch2doh_lines}
\end{figure*}

\indent\indent
Two lines of CH$_2$DOH have been detected towards L1544, $2_{0,2}-1_{0,1}$ 
and $3_{0,3}-2_{0,2}$, at~98.4 and~134.1\,GHz, respectively.
Both are $a$-type transitions belonging to the lower $e_0$ torsional state 
(see Table~\ref{tab:ch3oh_spec}).
The observations are shown in Figure~\ref{fig:ch2doh_lines}, and the results of the CLASS 
Gaussian fits are reported in the last two rows of Table~\ref{tab:ch3oh_fits}.
Interestingly, both CH$_2$DOH lines peak at a slightly bluer velocity compared to the ones 
of the parent species, which in turn are consistent with the systemic velocity of L1544
(7.2\,km\,s$^{-1}$, \citealt{Caselli-ApJ02-L1544k})\@.
The difference, \textit{ca.}\ 0.3\,km\,s$^{-1}$, is small but significant given the fit 
statistical errors ($<0.003$\,km\,s$^{-1}$ for most lines) and the spectral channel 
spacing of the observations ($0.11-0.15$\,km\,s$^{-1}$)\@.
It is also unlikely to be due to uncertainties in the rest frequencies because they are 
predicted to be as small as $\sim 0.005$\,km\,s$^{-1}$.
The slight discrepancy is thus suggestive of a complex cloud dynamics, and it suggests that 
deuterated methanol traces a more confined (maybe inner) region with respect to its parent 
species.
To confirm this, a map of CH$_2$DOH is required.

The CH$_2$DOH column density can be calculated from each transition assuming LTE 
conditions and optically thin emission once a suitable constraint for the excitation 
temperature, $T_\mrm{ex}$, is available.
It holds that
\begin{equation} \label{eq:N}
 N = \frac{8\pi\nu^3}{c^3} \frac{Q(T_\mrm{ex})}{g_u A_{ul}} 
     \frac{\ee^{E_u/kT_\mrm{ex}}}{\ee^{h\nu/kT_\mrm{ex}} - 1}
     \left[J_\nu(T_\mrm{ex}) - J_\nu(T_\mrm{bg})\right]^{-1}
     \int T_\mrm{mb}\diff v \,.
 \end{equation}
Given the uncertainty of the excitation temperature determined for the parent species,
the column density of the deuterated variant is derived through Eq.~\eqref{eq:N} assuming 
$T_\mrm{ex}$ ranging between~5 and~8\,K. 
The results are collected in Table~\ref{tab:ch2doh}. 
\begin{table}[ht]
  \caption[]{CH$_2$DOH column densities determined in L1544 assuming optically thin
             emission and excitation temperature in the 5--8\,K interval.}
  \label{tab:ch2doh}
  \vspace{1ex}
  \centering\small
  \begin{tabular}{l ccc}
    \hline\hline \\[-1ex]
           &  \mcl{3}{c}{$N$ / $10^{12}$cm$^{-2}$} \\
           &  \cline{1-3} \\[-2ex]
           & $T_\mrm{ex} = 5$\,K & $T_\mrm{ex} = 6.5$\,K & $T_\mrm{ex} = 8$\,K \\[0.5ex]
    \hline \noalign{\smallskip}
    $Q(T)^a$                  &  9.603 & 15.251 & 22.447 \\[0.5ex]
    $2_{0,2}-1_{0,1}$ ($e_0$) &  2.11  &  2.14  &  2.43  \\[0.5ex]
    $3_{0,3}-2_{0,2}$ ($e_0$) &  2.96  &  2.30  &  2.22  \\[0.5ex]
    \hline \\[-1ex] 
    \mcl{4}{l}{$^a$ Rotational partition function.}
  \end{tabular}
\end{table}

The rotational partition function, $Q(T)$ (whose values are also reported in 
Table~\ref{tab:ch2doh}) refers to the entire CH$_2$DOH population (i.e., it includes 
$e_0$, $e_1$, and $o_1$ levels) and was calculated by summing over rotational levels 
using the spectroscopic data of \citet{Pears-JMS12-CH3OH}.
The error bar of the CH$_2$DOH column density is estimated adding in quadrature the 
calibration error ($\sim$15\%) and the maximum dispersion due to the uncertainty in 
$T_\mrm{ex}$.
In this way we obtain \mbox{$N$(CH$_2$DOH)}$\:= (2.4\pm 0.9)\times 10^{12}$\,cm$^{-2}$\@.

We also performed sensitive observations at the frequency of the 
$1_{1,0}-1_{0,1}$ ($A^--A^+$) line of the singly deuterated CH$_3$OD isotopologue, but
they resulted in a non-detection.
The achieved $3\sigma$ sensitivity was 7\,mK\,km\,s$^{-1}$ (assuming a line width 
similar to that found for CH$_2$DOH) thus, with $T_\mrm{ex}$ constrained in the 5--8\,K 
interval, we derived a $3\sigma$ upper limit of $2.4\times 10^{11}$\,cm$^{-2}$ for the 
beam-averaged CH$_3$OD  column density (see next section for  a possible explanation of 
this non-detection).

\section{Discussion and conclusion}
\indent\indent
We have detected two lines of CH$_2$DOH in L1544 and carried out an accurate evaluation
of the methanol deuteration in a cold pre-stellar gas.
From the beam-averaged column densities computed in \S~\ref{sec:nlte} and~\ref{sec:ch2doh},
we obtain a fractionation ratio [CH$_2$DOH]/[CH$_3$OH] = $0.10\pm 0.03$\@, a value 
significantly lower than the ones measured in low-mass Class~0 protostars 
\citep[0.4--0.6,][]{Parise-AA06-GrainChem}, and lower than the deuterium fraction measured 
in molecules such as N$_2$H$^+$, NH$_3$, and H$_2$CO 
\citep{Caselli-ApJ02-L1544i,Bacmann-ApJ03-COdep,Roueff-AA05-NH3,Crapsi-AA07-L1544}.
However, it should be noted that, unlike methanol, all these molecules can also form in 
the gas phase, so that only our results reflect the surface chemistry activity directly.

State-of-the-art gas--grain chemical models of deuterium chemistry have been recently 
published by \citet{Aikawa-ApJ12-Dchem} and \citet{Taquet-ApJ12-Deut}.
These studies follow the molecular evolution and the D-fractionation as star formation 
proceeds from the pre-collapse phase to a proto-stellar core.
Both models are able to reproduce the high [CH$_2$DOH]/[CH$_3$OH] ratio observed in 
Class~0 protostars if D and H abstraction and substitution are included in the chemical
network.
At densities of $5\times 10^6$\,cm$^{-3}$ and $T = 10$\,K, the observed deuterium 
enhancement is reached in the molecular ice after $10^4-10^5$\,yr (relatively faster, 
$\sim 5\times 10^3$\,yr, following \citeauthor{Taquet-ApJ12-Deut} calculations).
On the other hand, a recent study of the gas-phase D-fractionation process 
\citep{Kong-Arxiv2013} indicates that the gas within the inner core of L1544 has a 
deuteration age (estimated through N$_2$H$^+$) of five to eight times the free-fall time scale, 
which is of the order of $10^4$\,yr.
This suggests that the deuterium reservoir frozen on grains should be fully developed,
at least in the central region of the core.

However, it should be noted that the column densities have been derived from observations 
towards the core centre, where CH$_3$OH suffers a considerable depletion 
(see Figure~\ref{fig:ch3oh_maps}).
Thus, the present observations may mostly be sensitive to the outer parts of the core, 
where the deuterium fractionation is lower than in the region traced by N$_2$H$^+$, a
molecular ion not significantly depleted at high gas densities 
\citep[e.g.,][]{Bizz-AA13-L1544}.
The observed methanol deuteration is intermediate between the N$_2$D$^+$/N$_2$H$^+$ (0.2) and 
DCO$^+$/HCO$^+$ (0.04) found by \citet{Caselli-ApJ02-L1544i} in L1544\@. 
This gives the hint that methanol deuteration is indeed tracing the region where CO 
is freezing out (at densities of a few $10^4$\,cm$^{-3}$), where the D/H ratio 
is not high enough to reach D-fractions close to the ones found towards Class~0 sources
or the centre of L1544\@.
Inner regions are lost to view, since freeze-out is probably too efficient or because the 
product of deuteration remains on the grains, because centre grains are covered 
with N$_2$  towards the core (see, e.g., \citealt{Bertin-ApJ13-Uvdes}).
Also, the morphology of the CH$_2$DOH emission in L1544 is not known, so the low 
measured [CH$_2$DOH]/[CH$_3$OH] ratio might be produced by chemical inhomogeneities 
present in this pre-stellar core.
Further progress in this study requires sensitive, interferometric observations aimed at 
deriving a detailed view of the fractionation in the different regions of the source.

From our non-LTE modelling results, we may infer that gas-phase CH$_3$OH in L1544 is 
composed of an almost equi-molar mixture of $A$ and $E$ species, and the resulting $[E]/[A]$
ratio is $0.97\pm 0.26$.
Given the large associated uncertainty, we may conclude that this value is 
consistent with the one implied by the picture of methanol formation on thermalised 
dust grains, (e.g., $[E]/[A]\approx 0.7$ at 10\,K).

The singly-deuterated CH$_3$OD isotopologue was not detected by the present observations,
yielding an upper limit of $2.4\times 10^{11}$\,cm$^{-2}$ for its column density.
The corresponding ratio between the singly-deuterated forms of methanol is 
[CH$_2$DOH]/[CH$_3$OD] $\geq 10$, higher than the nominal value of~3 that is predicted 
on a statistical basis assuming that D atoms are randomly distributed
in the methanol isotopologues \citep{Rodgers-PSS02-DD}.
Our finding confirms the trend reported by \citet{Bacman-MSL07-CH2DOH}, and it agrees
with the results of Class~0 protostars, where CH$_3$OD appears to be under-abundant with 
respect to the other methanol isotopic species ([CH$_2$DOH]/[CH$_3$OD] $\sim 14-20$, 
\citealt{Parise-AA06-GrainChem}).
Also, large [CH$_2$DOH]/[CH$_3$OD] ratios have been found by \citet{Rataj-AA11-CH3OH} in
a sample of low- to high-mass protostars.

The explanation of these ``anomalous'' ratios is still a challenge for the gas--grain 
chemical models, because H and D abstraction and substitution reactions --- whose rates on ices 
are not very well constrained --- are crucial to reproducing the observed abundances of the 
various methanol deuterated forms \citep[e.g.,][]{Taquet-ApJ12-Deut}. 
However, a laboratory study of low-temperature formaldehyde hydrogenation 
\citep{Hidaka-JPcs09-Hices} shows that the formation of CH$_2$DOH in ice dominates CH$_3$OD, 
owing to the higher velocity of the H--D substitution process compared to the D-atom addition.

The CH$_3$OH emission has a highly asymmetric annular distribution surrounding the dust 
peak, where CO is mainly frozen onto dust grains.
Methanol is expected to form via successive hydrogenation of CO on the surface of dust 
grains and then partially released in the gas phase upon formation (i.e., part of the 
formation energy is used to evaporate, in a process called reactive desorption; 
\citealt{Garrod-FD06-CH3OH}) and/or upon photo-desorption by UV photons produced by 
cosmic-ray impacts with H$_2$ molecules \citep{Prasad-ApJ83-UV}.
Evaporated methanol will then freeze-out onto dust grains in a time scale inversely 
proportional to the gas number density ($\sim 10^9/n_\mrm{H}$\,yr; \citealt{vDish-93-book}). 
Thus, the gas phase abundance of methanol is expected to decrease towards the centre of the 
core, where the density (and the freeze-out rate) is higher and where the outer layers of 
ice mantles may be rich in N$_2$, preventing hydrogenation of the underlying CO-rich 
layers \citep{Bertin-ApJ13-Uvdes,Vasy-ApJ13-grainsII}.

%

\begin{acknowledgement}
We are grateful to the IRAM $30\mut{m}$ staff for their support during the observations.
L.B. and E.L. gratefully acknowledge support from the Science and Technology Foundation 
(FCT, Portugal) through the Fellowships SFRH/BPD/62966/2009 and SFRH/BPD/71278/2010\@.
PC acknowledges the financial support of the European Research Council (ERC; project 
PALs~320620).
L.B. also acknowledges travel support to Pico Veleta from TNA Radio Net project funded 
by the European Commission within the FP7~Programme.
\end{acknowledgement}


\end{document}